# Leakage Tests of the Stainless Steel Vessels of the Antineutrino Detectors in the Daya Bay Reactor Neutrino Experiment

Xiaohui Chen(陈晓辉)[1], Xiaolan Luo(罗小兰)[1], Yuekun Heng(衡月昆)[1], Lingshu Wang(王灵淑)[1], Xiao Tang(唐晓)[1], Xiaoyan Ma(马晓彦)[1], Honglin Zhuang(庄红林)[1], Henry Band[2], Jeff Cherwinka[2], Qiang Xiao(肖强)[2], Karsten M.Heeger[2]

(Institute of High Energy Physics Beijing, 100049, China)

(University of Wisconsin, Madison, Wisconsin, 53706, USA)

**Abstract** The antineutrino detectors in the Daya Bay reactor neutrino experiment are liquid scintillator detectors designed to detect low energy particles from antineutrino interactions with high efficiency and low backgrounds. Since the antineutrino detector will be installed in a water Cherenkov cosmic ray veto detector and will run for 3 to 5 years, ensuring water tightness is critical to the successful operation of the antineutrino detectors. We choose a special method to seal the detector. Three leak checking methods have been employed to ensure the seal quality. This paper will describe the sealing method and leak testing results.

[1] Email: xhchen@ihep.ac.cn

# 1. Introduction

The Daya Bay Reactor Neutrino Experiment aims to measure $\sin^2 2\theta_{13}$ to a sensitivity of 0.01 or better at 90% confidence level [1]. The high sensitivity of the experiment demands good performances from the antineutrino detector (AD) which is constructed in three regions (Fig1). The inner-most zone (region I) is the Gd-doped liquid scintillator antineutrino target. The second zone (region II) is filled with normal liquid scintillator and serves as a γ-catcher to contain the energy of γ rays from neutron capture or positron annihilation. The outer-most zone (region III) contains mineral oil that shields external radiation from the fiducial volume.

The ADs will be installed in a pool (Fig. 2) filled with ultrapure water and operated under water for about 5 years. The quality of the AD seals is critical to the performance of the experiment. We need to ensure that no water leaks into the AD from the water pool, and that the target liquids (Gadolinium-dopped liquid scintillator), gamma catcher (normal liquid scintillator) and oil buffer (mineral oil) do not leak into each other. We chose double O-rings to seal the AD (Fig:2). This method is also convenient for leak checking. The O-rings are custom-made with Viton (FLUORCARBON RUBB/ER V7500A). With this method the AD can be sealed very well and can satisfy the design requirements. The criteria of the leak checking aims to guarantee, if there was a leak, that the leakage rate should be less then V/T, where the cumulative leaked volume V is not large enough to impact the physics performance of the AD, and Q is the experiment running time (5 years).

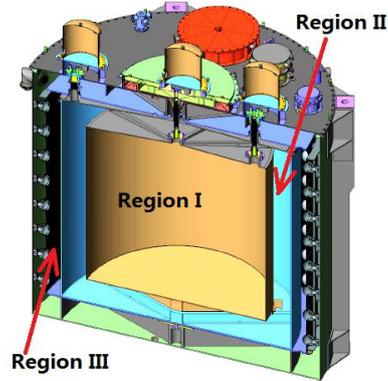

Fig1: Schematic top view of the different zones of the Daya Bay antineutrino detector

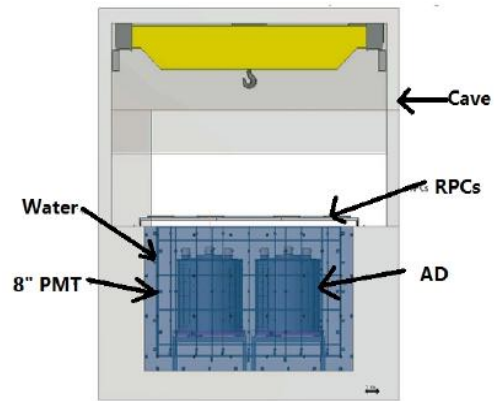

Fig2: Overview of the Daya Bay detectors in the experimental hall [1], ADs are installed side-by-side in the pool filled with ultrapure water. The ADs are surrounded by 2.5m water to shield environmental and rock backgrounds. PMTs are installed in the water to make it a cosmic ray veto Cherenkov detector. Above the water pool there is a layer of resistive plate chambers (RPC), serving as another independent cosmic ray veto detector.

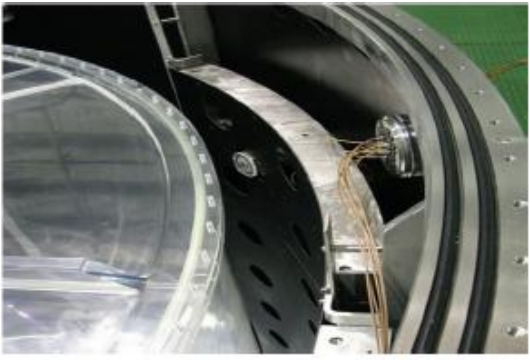

Fig 3: Picture of the double O-rings seal on the top flange of the main AD stainless tank.

## 2. The Methods and Principles of Leakage Test

Leakage tests have a very important role in vacuum technology. Gas leaks, also known as real leakage, mean that gas leaks from the higher pressure side to the side with lower pressure due to pin holes on the seal, etc. In this paper, we utilize gas leak check methods to ensure liquid seals. Specifically, the following methods have been considered [2]:

- Pressure Method: including water pressure method, bubbler method, and pressure change method.
- Vacuum Method: including static pressurizing method, Standard Vacuum Ionization Gauge method, and Helium Mass Spectrometry method.
- Back Pressure Method: combination of gas pressure method and vacuum method.

Due to special conditions during the installation of the detectors, we mainly use the gas pressure method(Fig:4), static pressurizing method(Fig:5), and method that simulates real conditions(water column method).

These methods will be discussed latter.

### 2.1 Gas Pressure Method

Gas Pressure Method is simple but very effective to get the leakage rate of the Gas out of the volume between the double O-rings (Fig: 4). This is different from the real situation when AD is in the water pool, because the gas within the volume leaks out of the double rings in two directions. In reality, the water will leak first through the outer O-ring and then through the inner rings to the AD if there was leakage

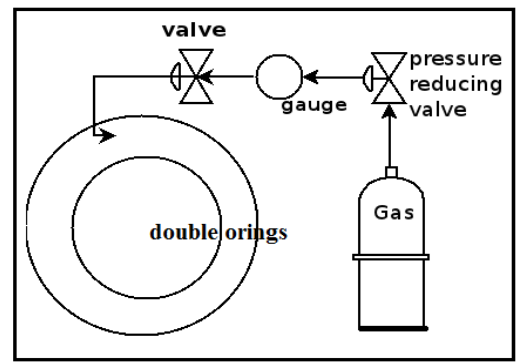

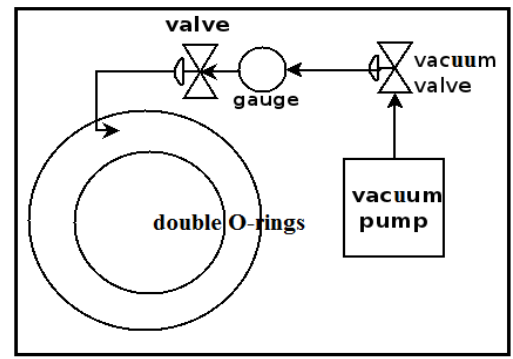

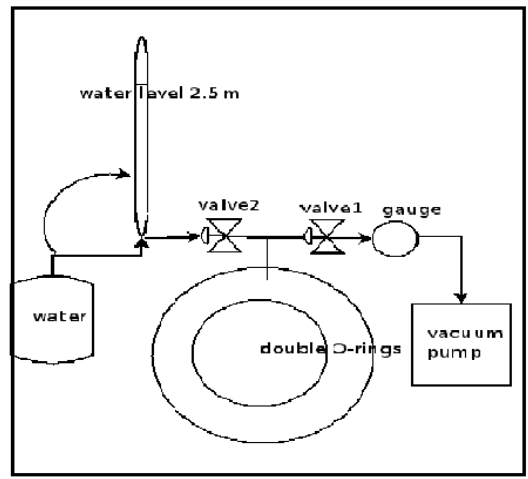

Fig.4: Diagram of pressure test, vacuum test and water column method

## 2.2 Vacuum Method

The gas pressure test gives the double-rings positive pressure, but the positive pressure will enhance the leakage. In order to do cross-check we also tried another method: making a vacuum between the double-rings which is a negative pressure test on the seals (Fig: 4).

The method to calculate leakage rate is as the same as that in the gas pressure test.

## 2.3 Water Column Method

The viscosity coefficient is very different between liquid and gas (in general liquids' coefficient two orders of magnitude viscosity than that of the gas). If we want to know the amount of water leakage, we need to make conversions.

From Hagen-Poiseuille law [3], we can get that the leakage rate is proportional to the coefficient of viscosity.

$$Q = \frac{\pi d^4 * \Delta p}{128 \mu * L} \qquad (2)$$

Naively, one simply scales the viscosity between the liquid and gas to make the gas-to-water conversion. However, uncertainties are unavoidable for the following reasons. Liquid has surface tension which is negligible in gas. Therefore the critical point of leakage in liquid and gas is different. In another word, when there is a gas leak, there may be no leak for liquid. In order get more reliable results, we designed a special method to simulate the real conditions with the water pressure method, which we call water column method (Fig: 4).

For water column method, the most difficult thing is how to fill the space between the double O-rings. First the space between the O-rings is pumped down to near vacuum (~mTorr). Then the valve to the vacuum pump is closed and a valve connected to a water source is opened. Water is then sucked into the space between the O-rings.

## 2.4 Leakage Rate

Assume the volume of the system we want to check is V, and the pressure varies with time as $dp/dt$, then formula calculating the leakage rate is:

$$Q = \frac{dp}{dt} * v \qquad (1)$$

Many factors will influence the leakage rate. The most important factor is the surface roughness and cleanliness of the grooves. The quality of the O-rings is also critical. The diameter of the O-rings should be even all along and the surface should be smooth. The details of the installation will influence the leakage rate. After each reinstallation we have to redo the leak checking, during the installation we need to apply the same torque to all bolts. This will make sure that the compression force to the O-rings is even. A proper torque (as specified by the O-ring manufacture) is also important.

## 3 Results

Our goal is that no water leaks into the AD, so the water column method will give a most clear and reliable result.

The leak checking of the first 2 ADs have been finished. For each AD, we mainly checked three parts: Two Overflow tanks and one main lid.

Fig5 show the leak test data using

different techniques on different double O-ring pairs. At the beginning we use the three methods to do the test and analyze the different of them, then in the latter test, we choose the most effective and simplest method (gas pressure test) to do the test.

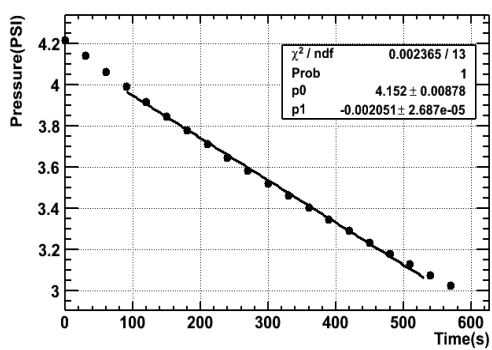

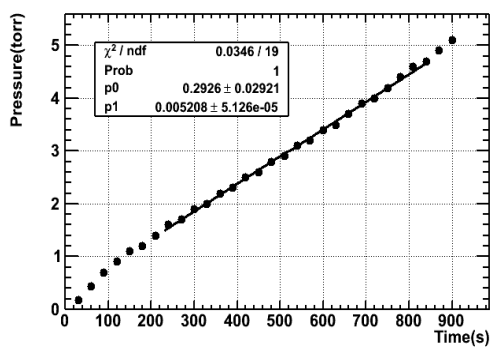

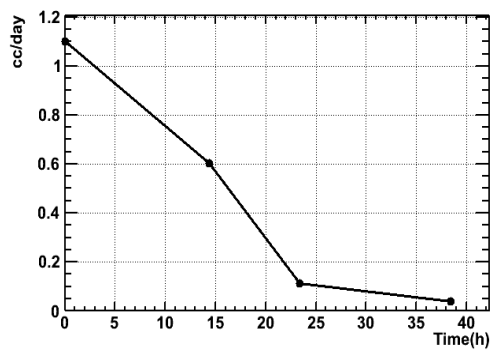

Fig5: The first figure is the result of the 2# Overflow Tank (AD1) using Gas Pressure Method: Q=1.56E-2cc/s, the second one is the result of the 1# Overflow Tank (AD2) using Vacuum Method: Q=7.6E-3cc/s, and the last one is the result of the 1# Overflow Tank (AD1) using water column method:
Q=0.04cc/day, the leakage rate is less than the criterion (0.06cc/s).

| AD parts | Method | result |
|---|---|---|
| Overflow Tank 1# | gas pressure | 1.56e-2 cc/s |
| | vacuum | 5.5e-4cc/s |
| | water column method | 4e-2cc/day |
| Overflow Tank 2# | gas pressure | 2.3e-3 cc/s |
| | vacuum | 1.69e-4 cc/s |
| | water column method | 3.3e-2cc/day |
| 5m SSV lid | gas pressure | 0.03 cc/s |
| | vacuum | 1.6E-3 cc/s |
| | water column method | No test |

Table 1: leak checking results of AD1

| AD parts | Method | result |
|---|---|---|
| Overflow Tank 1# | gas pressure | 5.2E-3cc/s |
| | vacuum | 7.6E-4 cc/sec |
| | water column method | 0.04cc/day |
| Overflow Tank 2# | gas pressure | 3.6E-3cc/s |
| | vacuum | 6.51E-4 cc/sec |
| | water column method | 0.03cc/day |
| 5m SSV lid | gas pressure | 5.2E-4cc/s |
| | vacuum | no test |
| | water column method | 0.07cc/day |

Table 2: leak checking results of AD2

From table1, and table2 we can see that the result of the vacuum method is better than that of the gas pressure method. That is because the negative pressure to double O-rings can reduce the leakage rate. The leakage rate of water is so small (<0.1cc/day) that we can assume that there is no leakage for water, there may be the evaporation of the water in the tube. When the results of the pumping vacuum method and the gas pressure method is good enough (Q<1E-3cc/s), there is no leakage detected using the water column method.

During latter test, if the results of the pumping vacuum method and the gas pressure method reach to 1E-4 cc/s or below, there is no need perform the leakage test with water column method.